\documentclass[12pt,preprint]{aastex}

\slugcomment{KUNS-1770, OU-TAP174}

\newcommand{\beq}{\begin{equation}}
\newcommand{\beqa}{\begin{eqnarray}}
\newcommand{\eeq}{\end{equation}}
\newcommand{\eeqa}{\end{eqnarray}}

\newcommand{\siml}{\lesssim}

\begin{document}


\title{  
X-ray Flashes from Off-axis Gamma-Ray Bursts
}


\author{Ryo Yamazaki\altaffilmark{1}}
\affil{Department of Physics, Kyoto University, Kyoto 606-8502, Japan}
\email{yamazaki@tap.scphys.kyoto-u.ac.jp}

\and

\author{Kunihito Ioka\altaffilmark{2}}
\affil{Department of Earth and Space Science, Osaka University, 
Toyonaka 560-0043, Japan}
\email{ioka@vega.ess.sci.osaka-u.ac.jp}

\and

\author{Takashi Nakamura\altaffilmark{3}}
\affil{Yukawa Institute for Theoretical Physics, Kyoto University,
Kyoto 606-8502, Japan}
\email{takashi@yukawa.kyoto-u.ac.jp}




\def\d{{\rm d}}
\def\p{\partial}
\def\w{\wedge}
\def\o{\otimes}
\def\f{\frac}
\def\tr{{\rm tr}}
\def\Half{\frac{1}{2}}
\def\half{{\scriptstyle \frac{1}{2}}}
\def\T{\tilde}
\def\RA{\rightarrow}
\def\N{\nonumber}
\def\n{\nabla}
\def\bb{\bibitem}
\def\BE{\begin{equation}}
\def\EE{\end{equation}}
\def\BEA{\begin{eqnarray}}
\def\EEA{\end{eqnarray}}
\def\L{\label}
\def\zero{{\scriptscriptstyle 0}}
\begin{abstract}
Ioka and Nakamura (2001) proposed a simple jet model 
that is compatible
with the peak luminosity-spectral lag relation, 
the peak luminosity-variability relation and 
various other relations in the Gamma-Ray Bursts. 
If the viewing angle is much larger than the collimation angle
of the jet in the model by Ioka and Nakamura, for appropriate model
parameters we obtain the observational characteristics of 
the X-ray flashes,
such as the peak flux ratio and the fluence ratio between 
the $\gamma$-ray ($50-300$ keV) and the X-ray band ($2-10$ keV), 
the X-ray photon index, the typical duration and the event rate
$\sim 100 \, {\rm yr}^{-1}$.
In our model, if the distance to the X-ray flashes is much
larger than $\sim 1$ Gpc (or $z \gtrsim 0.2$) they are too 
dim to be observed,
so the spatial distribution of the X-ray flashes should be
homogeneous and isotropic.
\end{abstract}




\section{INTRODUCTION}
Recently, a new class of X-ray transients has been recognized.
The Wide Field Cameras (WFCs) on the BeppoSAX in the X-ray range
$2-25\,{\rm keV}$ have detected some Fast X-ray Transients (FXTs)
with a duration less than $\sim 10^3\,{\rm s}$,
which are not triggered and not detected by 
the Gamma Ray Burst Monitor (GRBM) 
in the $\gamma$-ray range $40-700\,{\rm keV}$ 
(Heise et al. 2001; see also Strohmayer et al. 1998;
Gotthelf, Hamilton \& Helfand 1996;
Hamilton, Gotthlf \& Helfand 1996).
In Heise et al. (2001), these FXTs are defined 
as X-ray flashes (XRFs).
This definition of XRFs excludes the X-ray counterparts
of the typical Gamma-Ray Bursts (GRBs) including X-ray-rich GRBs.
Seventeen XRFs have been observed in the WFCs on the BeppoSAX in
about 5 yr, while 49 GRB counterparts have been observed in the
same period.  

XRFs have the following properties (Heise et al. 2001).
(i) The peak flux of the XRFs ranges between 
$10^{-8}$ and $10^{-7}\,{\rm erg}\ {\rm s}^{-1}\ {\rm cm}^{-2}$
(Fig.~2 of Heise et al. 2001).
The mean peak flux of the XRFs is 
about a factor of 3 smaller than that of the GRBs.
Nine out of 17 XRFs are detected in either the lowest or the
lowest two BATSE energy channels ($25-50$ and
$50-100\,{\rm keV}$; Kippen et al. 2001).
(ii) The ratio of the peak flux and the fluence in the X-ray range
($2-10\,{\rm keV}$) and the $\gamma$-ray range ($50-300\,{\rm keV}$)
for nine XRFs are shown in Fig.~3 of Heise et al. (2001).
The peak flux ratio extends up to a factor of 100,
and the fluence ratio extends up to a factor of 20.
(iii) The energy spectrum in the range $2-25\,{\rm keV}$ fits 
with a single power low  with the photon index between 1.2 and 3 
and the mean of about 2,
while the mean photon index of 36 GRBs in the same X-ray band 
is about 1, with the range between 0.5 and 3.
(iv) The duration of the XRFs ranges between 
$10\,{\rm s}$ and $200\,{\rm s}$,
which is the same order as that of the GRBs.
(v) The event rate of the XRFs is estimated 
as $\sim 100$ yr$^{-1}$ since the WFCs observed $\sim 3$ yr$^{-1}$
with the covering $40^\circ \times 40^\circ $ (full width to zero response).
(vi) The sky distribution is consistent with being isotropic.
The spatial distribution is consistent with being
homogeneous in Euclidean space since 
$\langle V/V_{max} \rangle=0.56 \pm 0.12$ (Heise et al. 2000;
Schmidt, Higdon \& Hueter 1988).

At present, the origin of the XRFs is not known.
Heise et al. (2001) have proposed that XRFs could be GRBs
at large redshift $z>5$,
when $\gamma$-rays would be shifted into the X-ray range.
However, as they have pointed out in their paper,
one cannot explain the duration distribution since
no time dilation due to cosmological expansion is observed.
There is also a possibility that the XRFs could be
dirty fireballs or failed GRBs 
(e.g., Dermer, Chiang \& B\"{o}ttcher1999; Heise et al. 2001;
Huang, Dai \& Lu 2002).

\begin{figure}
\plotone{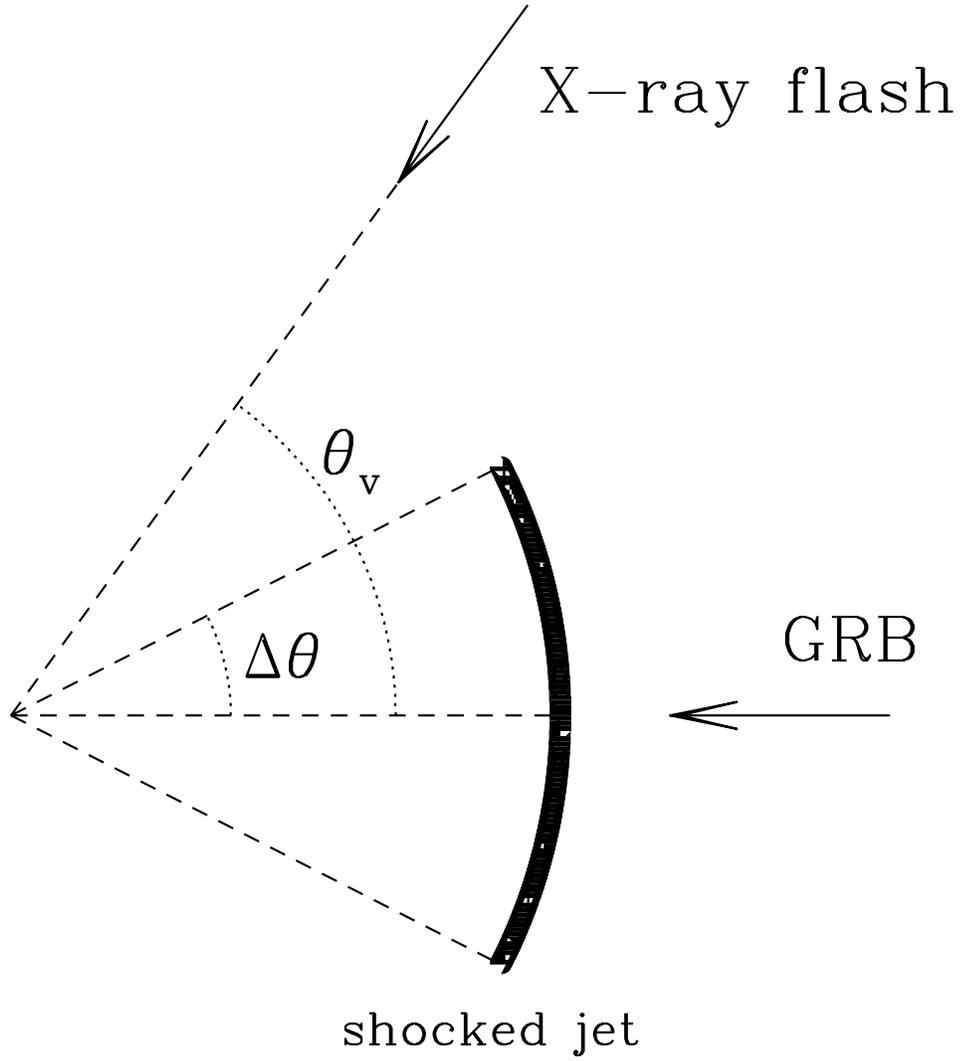}
\caption{
Our model is schematically shown.
The X-ray flashes are typical GRBs observed from
the large viewing angle.
}\label{fig:model}
\end{figure}

Ioka \& Nakamura (2001) have proposed that 
the XRFs could be GRBs
observed from the large viewing angle as
shown in Figure~\ref{fig:model} (see also Nakamura 2000).
They computed the kinematical dependence of 
the peak luminosity, the pulse width and the spectral lag of the peak
luminosity on the viewing angle $\theta_v$ of a jet.
For appropriate model parameters they obtained the 
peak luminosity-spectral lag relation similar to the observed one.
They suggested that the viewing angle of the jet
might cause various relations in GRBs
such as the peak luminosity-variability relation
and the luminosity-width relation.
Very recently several authors have also suggested that
the viewing angle is the key parameter to understand the
various properties of the GRBs
(Zhang \& M${\acute {\rm e}}$sz${\acute {\rm a}}$ros 2001;
Rossi, Lazzati \& Rees 2001; Salmonson \& Galama 2001).
In this circumstance, it is meaningful to study
the off-axis GRB model for the XRFs 
by Ioka \& Nakamura (2001) in more detail.

In this Letter, we will show that the GRBs
observed from the large viewing angle possesses
the above-mentioned properties (i)-(vi) of the XRFs.
In \S~\ref{sec:model}
we describe a simple jet model for the XRFs.
In \S~\ref{sec:ratio}
we consider the peak flux ratio and the fluence ratio
(property (ii)).
In \S~\ref{sec:index}
we consider the peak flux, the photon index and the event rate
(properties (i), (iii) and (v)).
\S~\ref{sec:dis} is devoted to discussion
(properties (iv) and (vi)).

\section{EMISSION MODEL OF X-RAY FLASHES}\label{sec:model}
We apply a simple jet model by Ioka \& Nakamura (2001) to the XRFs.
There are three timescales that determine 
the temporal pulse structure of the XRFs:
the hydrodynamic timescale $T_{dyn}$, 
the cooling timescale $T_{cool}$,
and the angular spreading timescale $T_{ang}$ 
(Kobayashi, Piran \& Sari 1997; Katz 1997; 
Fenimore, Madras \& Nayakshin 1996).
Since we consider that XRFs are the GRBs
observed from the large viewing angle,
we assume $T_{cool}\ll T_{dyn} \ll T_{ang}$
as in the case of GRBs (e.g., Piran 1999;
Sari, Narayan \& Piran 1996).
We adopt an instantaneous emission of an infinitesimally thin shell 
at $t=t_0$ and $r=r_0$.
Then the observed flux of a single pulse at the observed time $T$
is given by
\begin{eqnarray}
F_{\nu}(T)
={{2 c^2 \beta \gamma^4 A_0 (r_0/c\beta \gamma^2)}\over{D^2}}
{{\Delta \phi(T) f\left[\nu\gamma (1-\beta\cos\theta(T))\right]
}\over{\left[\gamma^2 (1-\beta\cos\theta(T))\right]^2}},
\label{eq:jetthin}
\end{eqnarray}
where $1-\beta\cos\theta(T)=({c\beta}/{r_0})(T-T_0)$
and $T_0=t_0-r_0/c\beta$.
The quantity $A_0$ determines the normalization of emissivity, and 
$f(\nu')$ represents the spectral shape
(for details, see Ioka \& Nakamura 2001, 
Granot, Piran \& Sari 1999, and Woods \& Loeb 1999).
Let the jet opening half-angle and the viewing angle be
$\Delta\theta$ and $\theta_v$, respectively (see Figure 1).
For $\Delta \theta>\theta_v ~{\rm and}~ 0<\theta(T)\le \Delta
\theta - \theta_v$, $\Delta \phi(T)=\pi$,
otherwise 
$\Delta \phi(T)=
\cos^{-1}\left[
{{\cos \Delta \theta - \cos \theta(T) \cos \theta_v}\over
{\sin \theta_v \sin \theta(T)}}\right]$.
For $\theta_v < \Delta \theta $, $\theta(T)$ varies from 0 to 
$\theta_v+\Delta \theta$ while from $\theta_v-\Delta \theta$ to
$\theta_v+\Delta\theta$ for $\theta_v > \Delta \theta $. 
In the latter case, 
$\Delta \phi(T)=0$ for $\theta(T)=\theta_v-\Delta \theta$.
A pulse starts at 
$T_{start}=T_0+({r_0}/{c\beta})
(1-\beta\cos(\max[0,\theta_v-\Delta\theta]))$ and ends at
$T_{end}=T_0+({r_0}/{c\beta})(1-\beta\cos(\theta_v+\Delta\theta))$.
The spectrum of the GRBs is well approximated by 
the Band spectrum (Band et al. 1993). 
In order to have a spectral shape similar to the Band spectrum,
we adopt the following form of the spectrum in the comoving frame,
\begin{eqnarray}
f(\nu')=\left({{\nu'}\over{\nu'_0}}\right)^{1+\alpha_B}
\left[1+\left({{\nu'}\over{\nu'_0}}\right)^{s}\right]^
{(\beta_B-\alpha_B)/s},
\label{eq:spectrum}
\end{eqnarray}
where $\alpha_B$ ($\beta_B$) is the low (high) energy power law index,
and $s$ describes the smoothness of the transition between
the high and low energy.  
In the GRBs, $\alpha_B\sim -1$ and $\beta_B\sim -3$ are typical values
(Preece et al. 2000).
Equations (\ref{eq:jetthin}) and (\ref{eq:spectrum})
are the basic equations to calculate the flux of a single pulse,
which depends on 10 parameters
for $\gamma \gg 1$, $\theta_v\ll 1$ and $\Delta \theta \ll 1$: 
$\gamma \nu_0'$, $\gamma \theta_v$, $\gamma \Delta \theta$, 
$r_0/c \beta \gamma^2$, $T_0$, $\alpha_B$, $\beta_B$, $s$, $D$, and 
$\gamma^4 A_0$.

In order to study the dependence on the viewing angle $\theta_v$,
we fix parameters as
$\gamma\Delta\theta=10$, $\alpha_B=-1$, 
$\gamma\nu'_0=300\,{\rm keV}$, $r_0/c\beta \gamma^2=10\,{\rm s}$ and $s=1$,
since typical GRBs have a break energy of $\sim 300\,{\rm keV}$
(Preece et al. 2000) and a pulse duration of $\sim 10\,{\rm s}$.
Other parameters, i.e.,
the viewing angle $\gamma\theta_v$, 
the high energy power law index $\beta_B$ and 
the distance $D$, are varied depending on circumstances.
We fix the amplitude $\gamma^4 A_0$ so that
the isotropic $\gamma$-ray energy 
$E_{iso}=4\pi D^2 S(20-2000\,{\rm keV})$
equals $10^{53}{\rm erg}$ when 
$\beta_B=-3.0$ 
and $\gamma\theta_v=0$.
Here $S(\nu_1-\nu_2)=
\int_{T_{start}}^{T_{end}}F(T;\nu_1-\nu_2)dT$
is the fluence in the energy range $\nu_1-\nu_2$
and $F(T;\nu_1-\nu_2)=\int_{\nu_1}^{\nu_2}
F_\nu(T)d\nu$ is the flux in the same energy range.
The result is
\begin{equation}
A_0=1.2\,{\rm erg}
\ {\rm cm}^{-2}
\ {\rm Hz}^{-1}
\left(\f{E_{iso}}{10^{53}{\rm erg}}\right)
\left(\f{r_0/c\beta\gamma^2}{10\,{\rm s}}\right)^{-2}
\left(\f{\gamma}{100}\right)^{-4}.
\label{Azero}
\end{equation}
Note that when we adopt $\gamma=100$,
the opening half-angle of the jet is similar to the observed one,
$\Delta \theta\sim 0.1$, and
the total energy corrected for geometry is comparable to 
the observed value,
$(\Delta\theta)^2E_{iso}\sim 10^{51}{\rm ergs}$
(Frail et al. 2001).

\section{PEAK FLUX RATIO AND FLUENCE RATIO}\label{sec:ratio}
In this section, we calculate the peak flux ratio
$R_{peak}=F_{peak}(2-10\,{\rm keV})/F_{peak}(50-300\,{\rm keV})$
and the fluence ratio
$R_{fluence}=S(2-10\,{\rm keV})/S(50-300\,{\rm keV})$
and compare the results with observations.

\begin{figure}
\plotone{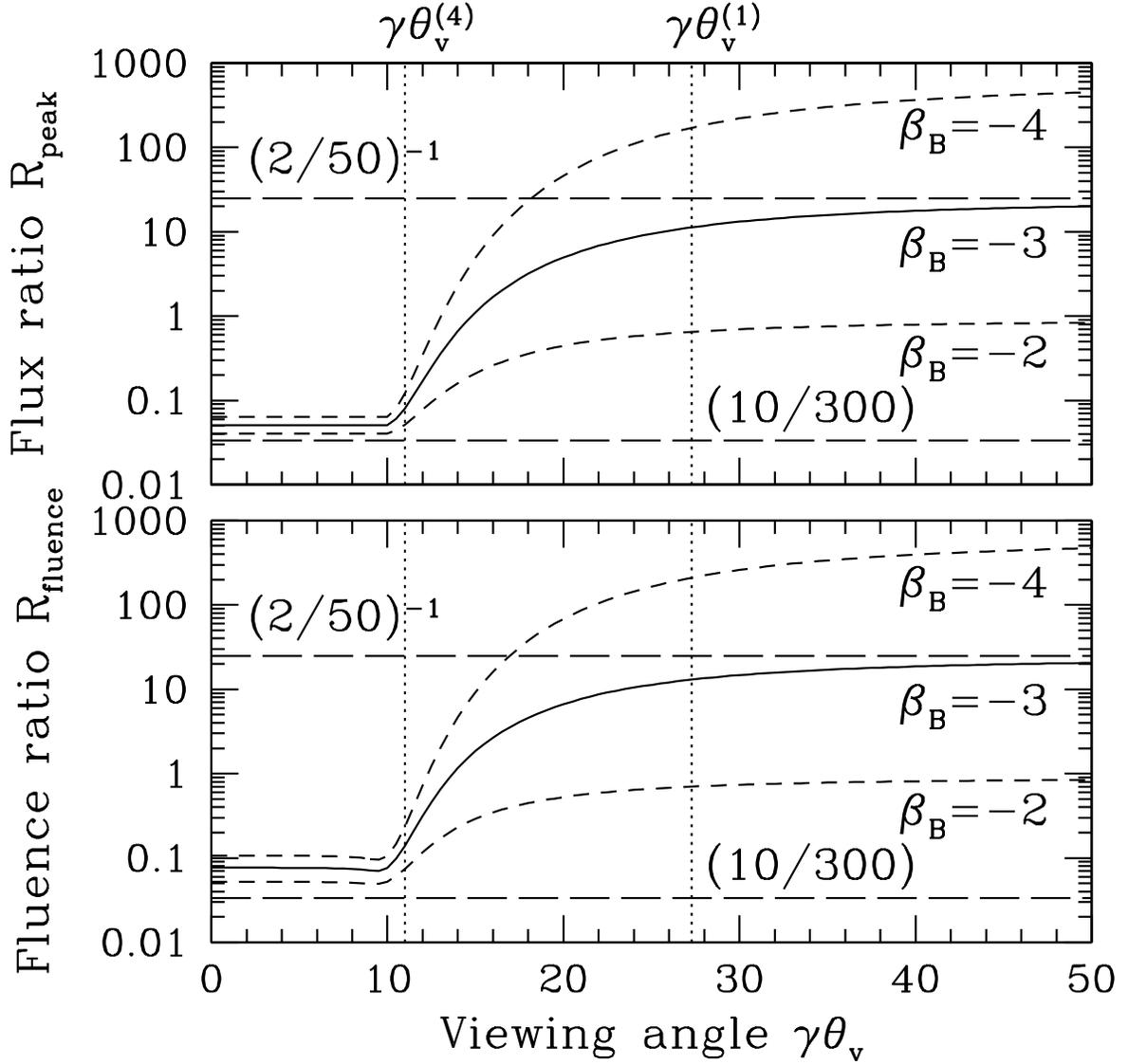}
\caption{
Peak flux ratio 
$R_{peak}=F_{peak}(2-10\,{\rm keV})/F_{peak}(50-300\,{\rm keV})$
(upper panel)
and fluence ratio
$R_{fluence}=S(2-10\,{\rm keV})/S(50-300\,{\rm keV})$ 
(lower panel)
as a function of the viewing angle $\gamma\theta_v$.
The solid curve shows the case $\beta_B=-3$,
and the dashed curves show the other cases, $\beta_B=-2$ and $\beta_B=-4$.
We adopt $\gamma\Delta\theta=10$, $\alpha_B=-1$,
$\gamma\nu'_0=300\,{\rm keV}$ and $s=1$.
The dotted line shows the viewing angle $\gamma \theta_v^{(1)}=27.3$ 
($\gamma \theta_v^{(4)}=11$) at which the maximum frequency $\nu_{max}$ 
equals the lowest (highest) observed energy, i.e.,
$2\,{\rm keV}$ ($300\,{\rm keV}$).
Here the maximum frequency $\nu_{max}$ means
the frequency at which most of the radiation energy is emitted.
At $\gamma \theta_v < \gamma \theta_v^{(4)}$
the ratios, $R_{peak}$ and $R_{fluence}$,
nearly equal $(\nu_2/\nu_4)^{2+\alpha_B}=(10/300)$,
and at $\gamma \theta_v > \gamma \theta_v^{(1)}$
the ratios, $R_{peak}$ and $R_{fluence}$,
nearly equal $(\nu_1/\nu_3)^{2+\beta_B}=(2/50)^{-1}$,
as shown by the long dashed lines.
}\label{fig:ratio}
\end{figure}

Figure~\ref{fig:ratio} shows the peak flux ratio $R_{peak}$
and the fluence ratio $R_{fluence}$ as a function of 
the viewing angle $\gamma\theta_v$.
When the viewing angle $\theta_v$ is larger than the 
opening half-angle $\Delta\theta$,
both the peak flux ratio $R_{peak}$ and 
the fluence ratio $R_{fluence}$ increase as 
the viewing angle $\gamma\theta_v$ increases.
The ratios, $R_{peak}$ and $R_{fluence}$, increase
as the high-energy index $\beta_B$ decreases.

We can understand this behavior as follows.
As shown in the Appendix,
the maximum frequency $\nu_{max}$ at which 
most of the radiation energy 
is emitted is estimated as $\nu_{max}\sim \nu_0'/\delta$,
where $\delta\equiv \gamma [1-\beta \cos (\theta_v-\Delta \theta)]
\simeq [1+\gamma^2 (\theta_v-\Delta \theta)^2]/2\gamma$
is the Doppler factor and $\theta_v > \Delta \theta$.
Thus the maximum frequency $\nu_{max}$ decreases 
as the viewing angle increases.
In the following, we consider two observation bands:
the lower energy band $\nu_1-\nu_2 \,{\rm keV}$,
and the higher energy band $\nu_3-\nu_4 \,{\rm keV}$.
The maximum frequency $\nu_{max}$ is larger than the highest 
observed energy $\nu_4$ 
($=300\,{\rm keV}$ in the present case) when 
$\gamma \theta_v<\gamma \theta_v^{(4)}\equiv\gamma \Delta \theta 
+\sqrt{2 \gamma \nu_0'/{\nu_4}-1}$.
In this case, we observe the low energy part
of the Band spectrum in equation (\ref{eq:spectrum}).
Since the low energy power law index is $\alpha_B=-1$,
the peak flux ratio $R_{peak}=F_{peak}(\nu_1-\nu_2\,{\rm keV})/
F_{peak}(\nu_3-\nu_4\,{\rm keV})$ and 
the fluence ratio $R_{fluence}=S(\nu_1-\nu_2\,{\rm keV})/
S(\nu_3-\nu_4\,{\rm keV})$ are given by
$R_{peak}\sim R_{fluence} \sim (\nu_2/\nu_4)^{2+\alpha_B}$,
where $\alpha_B>-2$.
Similarly, when the maximum frequency $\nu_{max}$ is 
smaller than the lowest observed energy $\nu_1=2\,{\rm keV}$,
i.e.,
$\gamma \theta_v > \gamma \theta_v^{(1)} \equiv\gamma\Delta\theta + 
\sqrt{2 \gamma \nu_0'/{\nu_1}-1}$,
the peak flux ratio and the fluence ratio are given by
$R_{peak} \sim R_{fluence} \sim (\nu_1/\nu_3)^{2+\beta_B}$,
where $\beta_B<-2$.

%

We compare Figure~\ref{fig:ratio} with observations.
Observed peak flux ratios extend up to a factor of 100 and
observed fluence ratios extend up to a factor of 20
(Fig.~3 of Heise et al. 2001).
One can see that 
when $\gamma \Delta \theta = 10\siml\gamma\theta_v\siml 
\gamma \theta_v^{(1)}\sim 
3 \gamma \Delta \theta$ and
$-4\siml\beta_B\siml -2$,
$R_{peak}$ and $R_{fluence}$ agree with 
the observational data.
Furthermore, Kippen et al. (2002) reported that $\nu_{max}$
ranges between about $2$ and $90 \,{\rm keV}$.
For our parameters, this can be reproduced if the viewing angle 
satisfies $\Delta\theta\lesssim\theta_v\lesssim\theta_v^{(1)}$.

\section{PEAK FLUX, PHOTON INDEX AND EVENT RATE}\label{sec:index}
We calculate the peak flux and the photon index in the energy
band $2-25\,{\rm keV}$ as a function of 
the viewing angle $\gamma\theta_v$,
and plot it in the peak flux $-$ photon index plane.
Figure 3 show the results for $\beta_B=-3$.
The distance is varied from $D=0.01$ Gpc to $D=2.1$ Gpc
for our parameters
\footnote{
When we consider the effect of cosmology 
($\Omega_M=0.3$, $\Omega_\Lambda=0.7$ and $h=0.7$), 
$D\sim2\,{\rm Gpc}$ corresponds to
$z\sim 0.4$. This does not affect our argument qualitatively,
but alters the quantitative results up to a factor of 2.}.
One can see that the photon index increases 
and the peak flux decreases as the viewing 
angle $\gamma\theta_v$ increases.

As discussed in \S~\ref{sec:ratio},
we observe the low- (high-) energy part
of the Band spectrum in equation (\ref{eq:spectrum})
when $\gamma \theta_v < \gamma \theta_v^{(4)}$
($\gamma \theta_v > \gamma \theta_v^{(1)}$),
where $\nu_4=25\,{\rm keV}$ and $\nu_1=2\,{\rm keV}$.
Therefore, the photon index in the energy range $2-25\,{\rm keV}$
is nearly equal to the low- (high-) energy spectral index 
$|\alpha_B|=1$ ($|\beta_B|=3$) when 
$\gamma \theta_v < \gamma \theta_v^{(4)}\simeq 14.8$
($\gamma \theta_v > \gamma \theta_v^{(1)}\simeq 27.3$).
With the analytical estimates in the Appendix,
we can also find that the peak flux $F_{peak}$ 
is approximately given by
\begin{eqnarray}
F_{peak} &\simeq& 4.3\times 10^{-6} {\rm erg}\,
{\rm s}^{-1}\,{\rm cm}^{-2}
\left(\frac{D}{1\,{\rm Gpc}}\right)^{-2}
\left[1+\gamma^2(\theta_v-\Delta \theta)^2\right]^{-2+\alpha_B}
\nonumber\\
&\times&\left(\frac{r_0/c\beta \gamma^2}{10\,{\rm s}}\right)
\left(\frac{\gamma \nu_0'}{300\,{\rm keV}}\right)^{-1-\alpha_B}
\left(\frac{\gamma^4 A_0}
{1.2\times 10^8\,{\rm erg}\ {\rm s}\ {\rm cm}^{-2}}\right),
\label{eq:fpeak}
\end{eqnarray}
when $\Delta \theta \siml \theta_v \siml \theta_v^{(4)}$
(In practice, eq.~[\ref{eq:fpeak}] can be applied to larger
viewing angles $\gamma\theta_v\lesssim30$.
We have confirmed that numerical results can be fitted within 5\% errors).
The peak flux $F_{peak}$ is smaller for larger viewing angle.
However, if the distances to such sources are small,
$F_{peak}$ may be comparable to that of the typical GRBs,
which have large distances and small viewing angles.

For comparison, we also plot the observed data
in the same figures (Fig.~2 of Heise et al. 2001).
One can see that the observed XRFs take place
within $\sim 2\,{\rm Gpc}$ and have
a viewing angle $\gamma \Delta \theta =10 \siml 
\gamma \theta_v \siml \gamma \theta_v^{(1)}\sim 3 
\gamma \Delta \theta$.

We roughly estimate the limits in flux sensitivity of the detectors.
On the right-hand side of the oblique dashed line,
the peak flux in the $\gamma$-ray band $F_{peak}(40-700\,{\rm keV})$
is larger than the limiting sensitivity of the GRBM
($\sim 10^{-8}\,{\rm erg}/{\rm s}\cdot{\rm cm}^2$),
and such events are observed as GRBs, not as XRFs.
The vertical dashed line represents the sensitivity of WFCs,
where we assume the integration time of $\sim{\rm ms}$.
Therefore, the observed data of XRFs sit in a fairly narrow region
surrounded by two dashed lines.

The distance to the farthest XRF $D_{XRF}$ gives
the observed event rate of the XRFs.
The observed event rate $R_{XRF}$ can be estimated as 
$R_{XRF}=r_{GRB}n_g (4\pi D_{XRF}^3/3)(f_{XRF}/f_{GRB})$,
where $r_{GRB}$ and $n_g$ are the event rate of the GRBs 
and the number density of galaxies, respectively.
The quantity
$f_{XRF}$ ($f_{GRB}$) is the solid angle subtended by the direction 
to which the source is observed as the XRF (GRB).
From previous discussions, one can find that
the emitting thin shell with opening half-angle $\Delta\theta$
is observed as the XRF (GRB) when the viewing angle
is within $\Delta\theta\siml\theta_v\siml \theta_v^{(1)}
\sim 3\Delta\theta$ ($0\siml\theta_v\siml \Delta\theta$).
Therefore the ratio of each solid angle is estimated as
$f_{XRF}/f_{GRB}\sim (3^2-1^2)/1^2=8$.
Using this value, we obtain 
\begin{eqnarray}
&&R_{XRF}\sim 10^2\,{\rm events}\ {\rm yr}^{-1}
\left(\f{r_{GRB}}{5\times 10^{-8}\,{\rm events}
\ {\rm yr}^{-1}\ {\rm galaxy}^{-1}}\right)
\left(\f{D_{XRF}}{2\,{\rm Gpc}}\right)^3\N\\
&&\times
\left(\f{n_g}{10^{-2}\,{\rm galaxies}\ {\rm Mpc}^{-3}}\right)
\left(\f{f_{XRF}/f_{GRB}}{8}\right),
\end{eqnarray}
which is comparable to the observation.

\section{DISCUSSION}\label{sec:dis}
We have shown that the observed data of the XRFs
can be reproduced by a simple jet model of the GRBs.
This suggests that the XRFs are identical to GRBs.
We may say that in the context of our model,
nearby GRBs are observed as XRFs when we see them
from the off-axis viewing angle.
If the distance to the XRFs is much larger than a few Gpc,
they cannot be observed since the observed flux is low.
This is consistent with the observed value of 
$\langle V/V_{max} \rangle\sim 0.5$
since the nearby sources distribute homogeneously in Euclidean space.

Our view of the XRFs is different from that of Heise et al. (2001).
They have proposed that XRFs could be GRBs at large
redshift $z>5$,
when $\gamma$-rays would be shifted into the X-ray range.
However, 
the observed total duration $T_{90}^{(obs)}$ cannot be explained.
In our model, $\gamma$-rays are shifted into the X-ray range 
by the relativistic beaming effect.
The total duration is equal to 
the lifetime of the central engine
and thus does not depend on the viewing angle $\theta_v$.
Hence the total duration of the XRFs may be
similar to that of the GRBs in our model.

We can calculate $T_{90}$, the observed duration of 
a single pulse in the X-ray band ($2-25\,{\rm keV}$).
When the viewing angle ranges from $\gamma \theta_v=10$ 
to $\gamma \theta_v=30$, the pulse duration is about
$T_{90}\sim 30-3000\,{\rm s}\,
[(r_0/c\beta\gamma^2)/10\,{\rm s}]$.
This value is comparable but a little bit inconsistent
with the observation 
since the observed pulse duration $T_{90}$, 
which is the order of the angular spreading timescale,
should be less than the total duration 
$T_{90}^{(obs)}\sim 10-200 \,{\rm sec}$,
which is the time interval between the first and the last emission.
This contradiction can be resolved as follows.
So far, we have assumed the isotropic energy of the 
instantaneous emission $E_{iso}\sim10^{53}\,{\rm erg}$ and
the time unit
$r_0/c\beta\gamma^2\sim 10\,{\rm s}$.
The effect of changing the values of these two parameters
appears only in the flux normalization 
$(\gamma^4 A_0)(r_0/c\beta \gamma^2)$ in equation (\ref{eq:jetthin}).
However, one can see that from equation (\ref{Azero}),
if one rescales these parameters as 
$E_{iso}\rightarrow E'_{iso}=10^{53}N^{-1}\,{\rm erg}$ and
$r_0/c\beta\gamma^2\rightarrow (r_0/c\beta\gamma^2)'=
10N^{-1}\,{\rm s}$,
the flux normalization factor is invariant 
$(\gamma^4 A_0)(r_0/c\beta \gamma^2)=[(\gamma^4 A_0)(r_0/c\beta \gamma^2)]'$,
which implies that the result is unchanged.
The value of $N$ is the number of instantaneous emissions,
since we fix the total emission energy as
$E_{iso}^{(tot)}=10^{53}\,{\rm ergs}$.
If we adopt $N\gtrsim 15$, 
$T_{90}$ of each emission can be less than $T_{90}^{(obs)}$.

Ioka \& Nakamura (2001) showed that the variability of GRBs is small 
for large viewing angle.
In addition, our model predicts that the number of pulses of
the XRFs is smaller than that of typical GRBs.
This can be expected from the following discussions.
In this Letter, we consider the time-averaged emissions,
which means that successive emissions from multiple subjets 
with the opening half-angle 
$\Delta\theta^{(j)}\sim\gamma^{-1}\sim\Delta\theta/10$
are approximated by one spontaneous emission
caused by a single jet with the viewing angle $\theta_v$
and the opening half-angle $\Delta\theta$.
Let the viewing angle of each subjet to be $\theta_v^{(j)}$.
The observed flux (or fluence) in the X-ray band due to
the subjets with $\theta_v^{(j)}\sim\theta_v+\Delta\theta$ are
much smaller than that with $\theta_v^{(j)}\sim\theta_v-\Delta\theta$,
and hence negligible.
We have confirmed this in the practical calculation.
If $\theta_v\gtrsim\Delta\theta$, the emissions of subjets with
$\theta_v^{(j)}\sim\theta_v-\Delta\theta$ dominates,
while if $\theta_v\sim0$, in the GRB case, the emissions from
almost all subjets may be detected.



\acknowledgments
We are grateful to the referee for useful comments.
This work was supported in part by
Grant-in-Aid for Scientific Research 
of the Japanese Ministry of Education, Culture, Sports, Science
and Technology  No.00660 (KI), 
No.11640274 (TN), No.09NP0801 (TN) and No.14047212 (TN).



\appendix

\section{ANALYTICAL ESTIMATES}\label{sec:app}
(I) $\nu_{max}$ and $(\nu S_{\nu})_{max}$:
In equation (\ref{eq:jetthin})
the typical value of $\theta(T)$ is $\sim (\theta_v-\Delta \theta)$ 
when $\theta_v>\Delta \theta$ 
since the flux peaks soon after the jet edge becomes visible.
Since the function $\nu' f(\nu')$ in equation (\ref{eq:spectrum})
takes a maximum at $\sim \nu_0'$, $\nu S_{\nu}$ takes a maximum at 
$\nu_{max}\sim \nu_0'/\delta \propto \delta^{-1}$
where $\delta\equiv \gamma [1-\beta \cos (\theta_v-\Delta \theta)]
\simeq [1+\gamma^2 (\theta_v-\Delta \theta)^2]/2\gamma$
and $S_{\nu}=\int_{T_{start}}^{T_{end}} F_{\nu}(T) dT$.
At $\nu_{max}$, $F_\nu$ in equation (\ref{eq:jetthin}) is proportional to
$\delta^{-2}$ so that we expect $(\nu S_{\nu})_{max} \propto \delta^{-3}$
(Ioka \& Nakamura 2001).
Note here that $\int \Delta \phi(T)dT$ depends on $\theta_v$ and $\delta$ 
very weakly.

(II) $T_{ang}$ and $\nu F_{\nu}^{peak}$:
The pulse duration $T_{ang}$ can be estimated 
as $T_{ang} \propto (T_{end}-T_{start}) \propto \theta_v^2 \propto \delta$ 
for $\theta_v \sim \Delta \theta$,
and $T_{ang} \propto (T_{end}-T_{start}) \propto \theta_v \propto
\delta^{1/2}$ for $\theta_v \gg \Delta \theta$.
The peak flux $F_{peak}$ can be estimated from the relation 
$F_{peak} T_{ang}
\sim S \propto \delta^{-1+\alpha_B} (\delta^{-1+\beta_B})$
when the maximum frequency $\nu_{max}$ 
is higher (lower) than the observed frequency.




\clearpage

\begin{figure}
\plotone{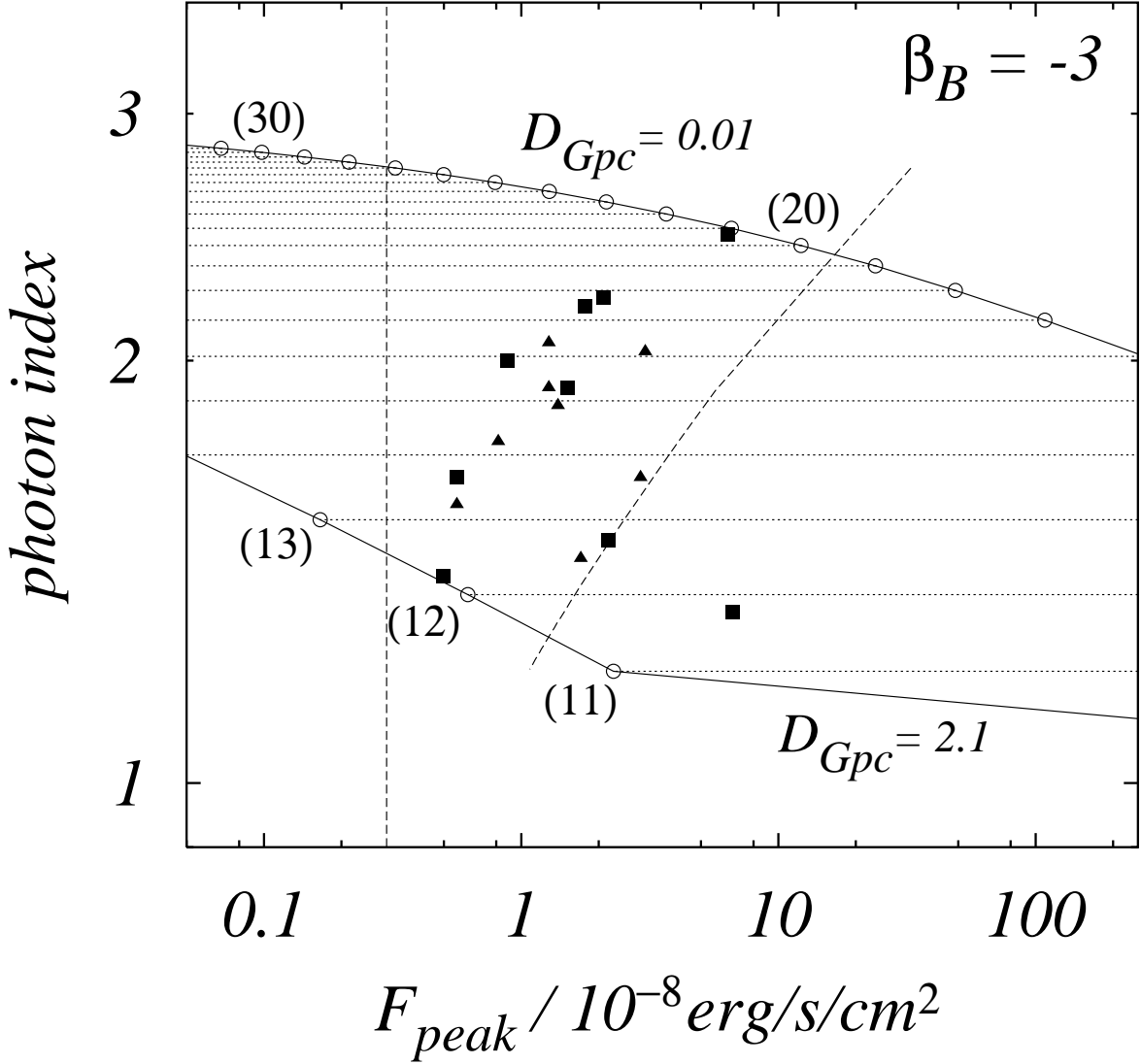}
\caption{
Photon index in the energy range $2-25\,{\rm keV}$
as a function of the peak flux in the same energy range
by varying the distance $D$.
We adopt $\gamma\Delta\theta=10$,
$\alpha_B=-1$, $\beta_B=-3$, $\gamma\nu'_0=300\,{\rm keV}$
and $s=1$.
The values of the viewing angle $\gamma\theta_v$ are
given in parenthesis.
The right-hand side (left side) of the two solid curves is 
$D=0.01\,{\rm Gpc}$ ($D=2.1\,{\rm Gpc}$).
Points that correspond to same values of $\gamma\theta_v$
but different $D$ are connected by horizontal dotted lines.
The observed data shown are from Heise et al. (2001).
Squares (triangles) are those which were (were not)
detected  by BATSE.
Two dashed lines represent observational bounds.
In the region to the left of the vertical dashed line,
the peak flux in the  X-ray band is smaller than the 
limiting sensitivity of WFCs 
(assuming the integration time of $\sim$ ms), 
and such events cannot be observed.
In the region to the right of the oblique dashed line,
the peak flux in the $\gamma$-ray band is larger than the limiting
sensitivity of GRBM 
($\sim 10^{-8}\,{\rm erg}/{\rm s}\cdot{\rm cm}^2$), 
and such events are observed as GRBs.
}
\end{figure}

\end{document}